\documentclass[reprint,aps,prl,noeprint,twocolumn,superscriptaddress,longbibliography]{revtex4-1}
\usepackage[latin9]{inputenc}
\usepackage{color}
\usepackage{multirow}
\usepackage{makecell}
\usepackage{mathtools}

\usepackage{amsmath}
\usepackage{amssymb}
\usepackage{stmaryrd}
\usepackage{graphicx}
\usepackage[unicode=true,
 bookmarks=true,bookmarksnumbered=false,bookmarksopen=false,
 breaklinks=false,pdfborder={0 0 1},backref=false,colorlinks=true]
 {hyperref}
\hypersetup{
 linkcolor=magenta, urlcolor=blue, citecolor=blue, pdfstartview={FitH}, hyperfootnotes=true, unicode=true}
 
\makeatletter
\@ifundefined{textcolor}{}
{%
 \definecolor{BLACK}{gray}{0}
 \definecolor{WHITE}{gray}{1}
 \definecolor{RED}{rgb}{1,0,0}
 \definecolor{GREEN}{rgb}{0,1,0}
 \definecolor{BLUE}{rgb}{0,0,1}
 \definecolor{CYAN}{cmyk}{1,0,0,0}
 \definecolor{MAGENTA}{cmyk}{0,1,0,0}
 \definecolor{YELLOW}{cmyk}{0,0,1,0}
}

\usepackage[normalem]{ulem}
\usepackage{cancel}

\usepackage{amsfonts}\usepackage{tabularx}\usepackage{dcolumn}\usepackage{bm}\usepackage{graphicx}\usepackage{epstopdf}

\hypersetup{urlcolor=blue}

\usepackage{tensor}
\usepackage{braket}

\usepackage[capitalise,compress]{cleveref}
\crefname{section}{Sec.}{Secs.}
\Crefname{section}{Section}{Sections}
\crefrangelabelformat{equation}{\textup{(#3#1#4)}--\textup{(#5#2#6)}}

\makeatother
\begin{document}
\title{Localization and criticality in antiblockaded 2D Rydberg atom arrays} 
\author{Fangli Liu}
\affiliation{Joint Quantum Institute and Joint Center for Quantum Information and Computer Science, NIST/University of Maryland, College Park, MD, 20742, USA}
\author{Zhi-Cheng Yang}
\affiliation{Joint Quantum Institute and Joint Center for Quantum Information and Computer Science, NIST/University of Maryland, College Park, MD, 20742, USA}
\author{Przemyslaw Bienias}
\affiliation{Joint Quantum Institute and Joint Center for Quantum Information and Computer Science, NIST/University of Maryland, College Park, MD, 20742, USA}
\author{Thomas Iadecola}
\affiliation{Department of Physics and Astronomy, Iowa State University, Ames, Iowa 50011, USA}
\author{Alexey V. Gorshkov}
\affiliation{Joint Quantum Institute and Joint Center for Quantum Information and Computer Science, NIST/University of Maryland, College Park, MD, 20742, USA}

\begin{abstract}
Controllable Rydberg atom arrays have provided new insights into fundamental properties of quantum matter
both in and out of equilibrium. In this work, we study the effect of experimentally relevant positional disorder on Rydberg atoms trapped in a 2D square lattice under anti-blockade (facilitation) conditions. We show that the facilitation conditions lead the connectivity graph of a particular subspace of the full Hilbert space to form a 2D Lieb lattice, which features a singular flat band. Remarkably, we find three distinct regimes as the disorder strength is varied: a critical regime, a delocalized but nonergodic regime, and a regime with a disorder-induced flat band. The
critical regime's existence depends crucially upon the singular flat band in our model, and is absent in any 1D array or ladder system. We propose to use quench dynamics to probe the three different regimes experimentally. 
\end{abstract}
\maketitle

Recently, programmable Rydberg quantum simulators have attracted significant interest because they can provide insights into quantum matter's fundamental properties. With the rapid development of quantum technologies, synthetic arrays of Rydberg atoms with individual control are already available in one~\cite{Endres16}, two~\cite{Barredo16, Guardado18}, and three dimensions~\cite{Barrodo18}. Recent experiments on 1D Rydberg atom arrays have shed light on various phenomena, including nonequilibrium quantum many-body dynamics~\cite{Bernien17}, the Kibble-Zurek mechanism~\cite{Keesling19}, and quantum many-body scars~\cite{Bernien17, Turner18}. The strong Rydberg-Rydberg interactions can also be used to realize quantum gates~\cite{Levine19}, making such systems promising platforms for quantum information processing~\cite{Levine18, Omran19}. 

Meanwhile, flat band systems 
are conceptually important in condensed matter physics 
and can harbor both nontrivial topological properties~\cite{Sun11, Tang11, Neupert11, Yao12} and strongly correlated phases arising from the enhanced interaction effects~\cite{Lieb89, Mielke_1991, Leykam18, Regnault11, Bergholtz13, Liu12, Norman13, Zhu16}.
Recent work on twisted graphene heterostructures and circuit quantum electrodynamics (QED) opens up new venues for flat bands, enabling, respectively, the study of correlated insulators and superconductivity~\cite{Cao18, Cao19, Yankowitz19, Po18} and of quantum systems in hyperbolic space~\cite{Kollar19, Kollar192}. One particular direction of interest concerns the effect of disorder on flat-band eigenstates. It has been shown that such flat bands, when coupled to disorder, can lead to 
critical and multifractal phenomena 
absent in conventional Anderson localization~\cite{Chalker10, Leykam13, Goda06, Wilson19, Wilson20, Fu20, Hunber10, Shukla18, Gneiting2018}.

In this work, we demonstrate that the physics of flat bands coupled to disorder can be naturally realized and probed using Rydberg atoms trapped in a
2D square lattice. 
We consider the so-called facilitation 
(anti-blockade) mechanism, where the excitation of a Rydberg atom is strongly enhanced in the vicinity of an already excited atom~\cite{Mattio15, Marcuzzi17, Ostmann19}. Under such conditions, the full Hilbert space can effectively split into subspaces separated from one another by large energy scales. We focus on
the manifold of states that can be created near-resonantly starting from a single Rydberg excitation.
Within this subspace, the system can effectively be described by a single particle hopping on a 2D Lieb lattice~\cite{Ostmann19}, which features a singular flat band in the clean limit. Although the Lieb lattice has been experimentally realized for 
photons~\cite{Mukher15, Vicencio15, Diebel16, Guzm14, Xia16},
atoms~\cite{Taie15, Baboux16},
and electrons~\cite{Slot17}, the effect of disorder 
on flat-band states has not yet been systematically studied.
We find that the interplay between positional disorder of Rydberg atom arrays 
and the synthetic flat-band states gives rise to a rich phase diagram, including a critical phase, a nonergodic extended phase, and a phase with a disorder-induced flat band. We show that these intriguing properties are essentially related to the singular flat band on the Lieb lattice and are absent in 1D and quasi-1D arrays.

\begin{figure}
  \centering\includegraphics[width=0.48\textwidth,height=8.8cm]{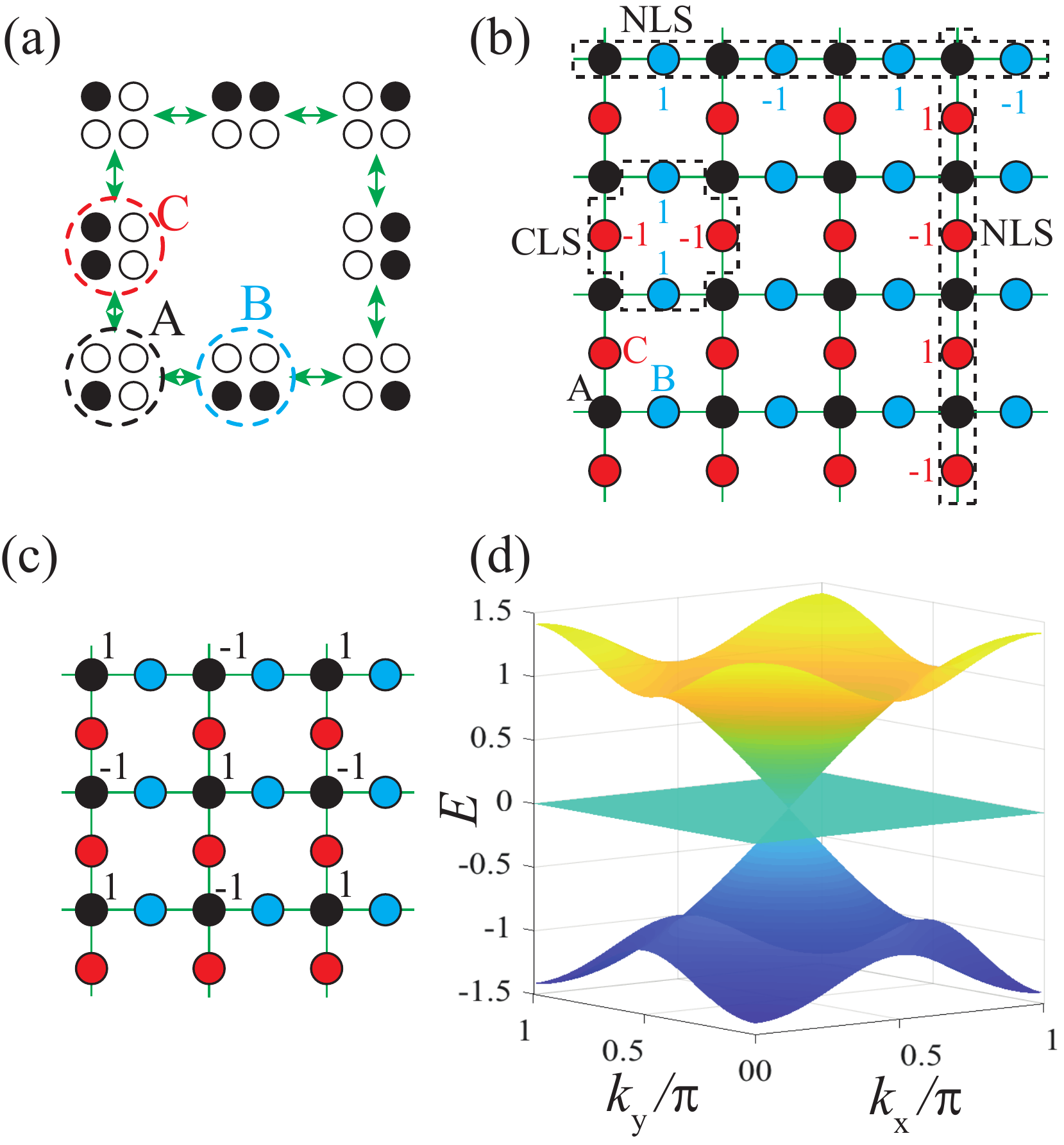}
  \caption{ (a) Under the anti-blockade conditions,  the connectivity graph of the subspace containing 
  single isolated Rydberg excitations and single nearest-neighbor pairs thereof
  maps to a 2D Lieb lattice shown in (b). The black and white dots indicate atoms in Rydberg and ground states,  respectively. Each unit cell of the Lieb lattice contains three 
  sites: $A$,  $B$, and $C$. (b) The flat-band eigenstates include compact localized states (CLSs), two non-contractible loop states (NLSs), and one non-compact state (NCS) shown in (c). The `$\pm 1$'  indicate the relative wavefunction amplitudes for these states. 
  (d) The band structure of the clean Lieb lattice, which contains two dispersive bands and one singular flat band. 
  }
\label{fig1}
\end{figure}

{\it Antiblockaded Rydberg atom array and mapping to Lieb lattice.---}We consider the following Hamiltonian describing interacting Rydberg atoms trapped in a 2D $L\times L$ square lattice with spacing $R_0$:
\begin{align}\label{eq:Rydberg_toy_confine}
H_{\rm Ryd}=&
\frac{\Omega}{2} \sum_{i}^{N}  \sigma^x_i - \Delta \sum_{i}^{N} n_i+  \frac{1}{2} \sum_{i \neq j}^{N} V(d_{ij})n_i n_{j},
\end{align}
where $i$ and $j$ run over sites of 
the square lattice [see Fig.~\ref{fig1}(a)], $\sigma^x_i= \ket{g_i}\!\bra{r_i}+\ket{r_i}\!\bra{g_i}$, $\ket{g_i}$ ($\ket{r_i}$) denotes the ground (Rydberg) state of the $i$-th atom, and $n_i=\ket{r_i}\bra{r_i}$. The parameters $\Omega$ (Rabi frequency) and $\Delta$ (detuning) characterize coherent driving fields, while $V(d_{ij}) \propto 1/d_{ij}^6$ quantifies the van-der-Waals interactions between atoms in Rydberg states at sites $i$ and $j$ (separated by distance $d_{ij}$).  The anti-blockade (facilitation) condition is obtained by setting $\Delta= V(R_0)$, so that an isolated excitation makes the excitation of its 
nearest neighbour resonant~\cite{Mattio15, Marcuzzi17, Ostmann19}. We 
work in the limit
$ |\Delta| \gg \Omega$   
where the un-facilitated excitations are sufficiently off-resonant. We additionally require $V(\sqrt{2} R_0) , V(2R_0) \gg \Omega$, so that a pair of neighbouring Rydberg excitations is unable to further facilitate the excitation of any neighbouring site. Hereafter we take $V(R_0) = 300 \Omega$.

Under these conditions, the 
Hilbert space effectively splits into subspaces that are separated by energy scales much larger than $\Omega$~\cite{Marcuzzi17}. Here we focus on the simplest nontrivial subspace, whose degrees of freedom are hardcore bosons consisting of either a single Rydberg excitation or a pair of neighbouring Rydberg excitations. 
One can readily see that the connectivity graph of states in this subspace forms a 2D Lieb lattice [see Figs.~\ref{fig1}(a)-(b)]. The Hamiltonian~(\ref{eq:Rydberg_toy_confine}) thus reduces to a single particle hopping 
on this lattice. The Lieb lattice contains three 
sites per unit cell, where the $A$
site corresponds to a single Rydberg excitation in the original model, while the $B$ and $C$  
sites correspond, respectively, to horizontal and vertical pairs of neighbouring Rydberg excitations [see Supplemental Material (SM) for more details~\cite{supp}].
 
{\it Flat band on the Lieb lattice.---} The 
single-particle hopping Hamiltonian on the Lieb lattice takes the form
\begin{equation}
H_{\rm Lieb} = \sum_{\langle i,j \rangle} \Omega \ c_i^\dagger c_j + {\rm H.c.},
\label{eq:lieb}
\end{equation}
where $\langle i,j\rangle$ denotes nearest-neighbor sites on the Lieb lattice, as shown in Fig.~\ref{fig1}(b).
The energy spectrum of Hamiltonian~(\ref{eq:lieb}) contains two dispersive bands $E_{\pm} ({\bm k})= \pm \Omega \sqrt{\cos^2{(k_x)} +\cos^2{(k_y)} }$ and one flat band $E=0$ [see Fig.~\ref{fig1}(d)].  The zero-energy flat band touches the two dispersive bands at $k_x=k_y=\pi/2$, leading to a three-fold degeneracy at this point. As shown in Refs.~\cite{Bergman08, Rhim19}, the band-touching in this model is in fact irremovable, which signals a singularity in the Bloch wavefunction. The $E=0$ band of Hamiltonian~(\ref{eq:lieb}) in this case is known as a \textit{singular} flat band. The singularity of the flat band has important consequences on the eigenstates within the band. Generically, the eigenstates of a flat band are localized in real space, hence the name compact localized states (CLSs) [see Fig.~\ref{fig1}(b) for the Lieb lattice]. When the flat band is non-singular, such CLSs form a complete basis of the flat band. By contrast, when the flat band is singular, the set of all CLSs is not linearly independent. 
For the Lieb lattice, there exist three additional extended eigenstates of the flat band: two non-contractible loop states (NLSs) [Fig.~\ref{fig1}(b)] and one non-compact state (NCS) [Fig.~\ref{fig1}(c)].

\begin{table}[]
\begin{tabular}{c|c|c|c}
\hline
                  &   Wavefunction    &    Support        &   Feature            \\ \hline \hline
\multirow{2}{*}{Regime I} & \multirow{2}{*}{critical, multifractal} & \multirow{2}{*}{$B$, $C$} & \multirow{2}{*}{original flat band} \\
                  &                   &                   &                 \\ \hline
\multirow{2}{*}{Regime II} & \multirow{2}{*}{multifractal} & \multirow{2}{*}{$A$, $B$, $C$} & \multirow{2}{*}{\makecell{hybridization with \\ dispersive bands}} \\
                  &                   &                   &                   \\ \hline
\multirow{2}{*}{Regime III} & \multirow{2}{*}{\makecell{localized ($|E|\gtrsim 0$), \\ multifractal ($E\approx 0$)}} & \multirow{2}{*}{$A$} & \multirow{2}{*}{\makecell{disorder-induced \\ flat band}} \\
                  &                   &                   &                   \\ \hline
\end{tabular}
\caption{Main features of three distinct localization regimes.}
\label{table1}
\end{table}

{\it Positional disorder.---}Small
deviations of atomic positions from the centers of the corresponding traps can significantly affect the atom-atom interaction. 
The thermal distribution of atomic positions 
can be described as a Gaussian with width $\sigma$ (measured in units of $R_0$)~\cite{Omran19, Marcuzzi17}. Ignoring atomic motion during the experiment (frozen-gas approximation)~\cite{Marcuzzi17}, such randomness enters Eq.~(\ref{eq:Rydberg_toy_confine}) via the interaction term: $V(R)= V(R_0+ \delta R) \approx  V(R_0)+ \delta V$, where $\delta V$ is a random time-independent shift potential caused by the displacement. 
This position-disordered interaction
manifests itself on 
the effective Lieb lattice
as random, but correlated, on-site potentials for the $B$ and $C$ sublattices. Since the  position disorder
only affects Rydberg-Rydberg interactions, the $A$ sublattice sites, which represent single Rydberg excitations, do not couple to disorder. Therefore, while the CLSs and NLSs are supported on $B$ and $C$ sublattices and hence are no longer exact eigenstates of the disordered Hamiltonian, the non-compact state in Fig.~\ref{fig1}(c) remains unaffected by disorder.

\begin{figure}
  \centering\includegraphics[width=0.5\textwidth,height=8.3cm]{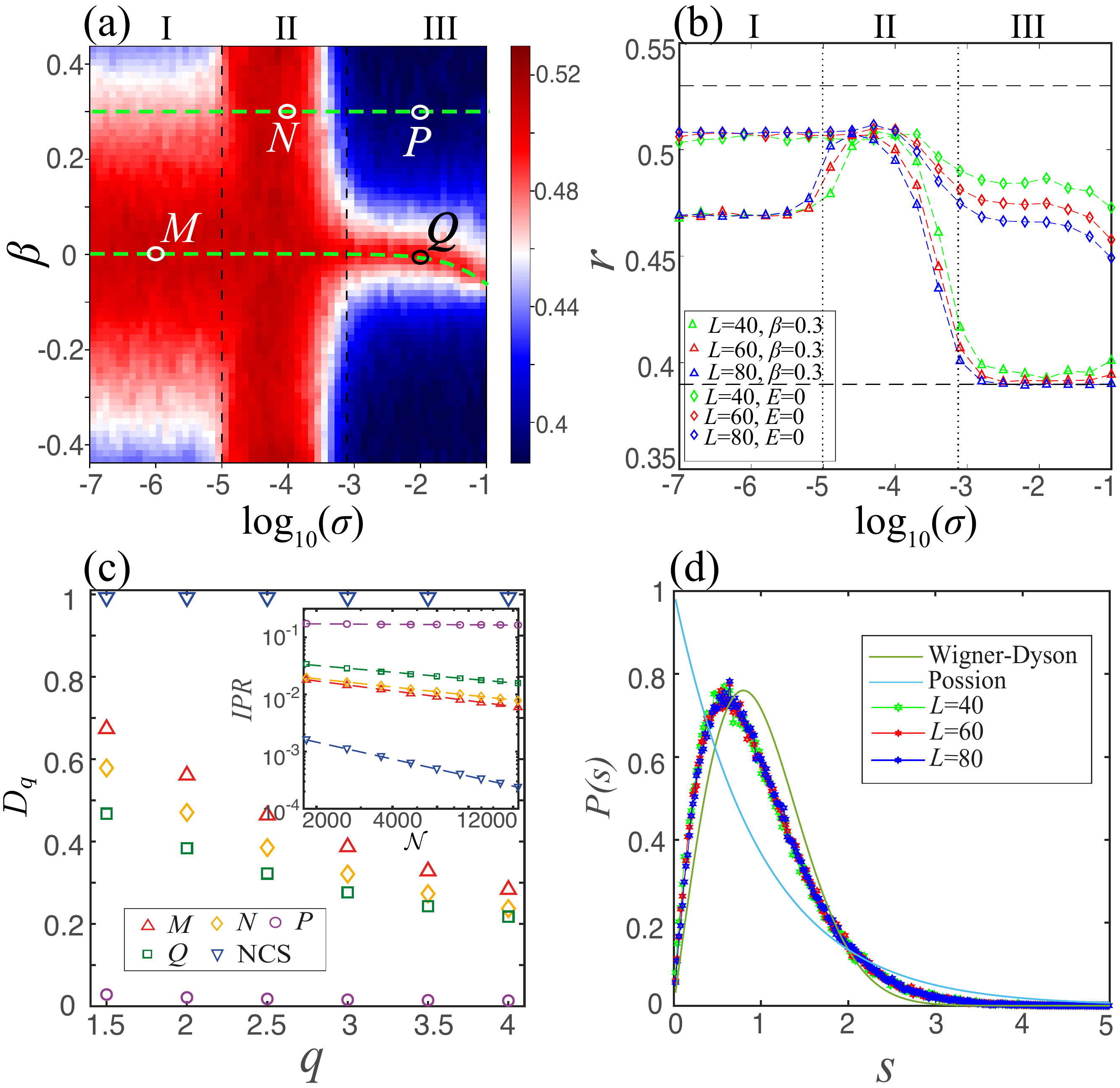}
  \caption{
  (a) Level-spacing ratio $r$ versus the rescaled eigenstate label $\beta$ and disorder strength $\sigma$. (b) $r$ as a function of disorder strength for two  cuts, shown by dashed lines in (a), along $E=0$ and $\beta=0.3$  for different system sizes. The error bars (not shown) are smaller than the plot markers. (c) Fractal dimension $D_q$ versus $q$, for states at representative points in (a): $M$ ($\beta=0, \log_{10}\sigma=-6$), $N$ ($\beta=0.3, \log_{10}\sigma=-4$), $P$ ($\beta=0, \log_{10}\sigma=-2$), $Q$ ($\beta=0.3, \log_{10}\sigma=-2$), as well as the non-compact zero-energy eigenstate (NCS) for arbitrary disorder strength. 
  Inset: scaling of IPR as a function of the Hilbert space dimension.   (d) Probability distribution of the unfolded level spacings $P(s)$ for states in regime I for different system sizes~\cite{dataverage}. 
  }
\label{fig2}
\end{figure}

 To study the effect of disorder on the singular flat band, we numerically diagonalize the Lieb lattice Hamiltonian~(\ref{eq:lieb}) in real space with positional disorder on an $L \times L$ square lattice.
  We focus on the middle one third of eigenstates in the spectrum, which corresponds to the flat-band states 
  in the clean limit.
  We rank-order the eigenstates according to their energies $E_i>E_{i-1}$ and introduce a rescaled label $\beta= \frac{i-\mathcal{N}/2 }{\mathcal{N}/3 } \in (-0.5, 0.5)$, where $\mathcal{N}$ is the Hilbert-space dimension and $i\in (\mathcal{N}/3, 2\mathcal{N}/3)$. We probe 
  ergodicity versus localization
  using the level-spacing ratio $r_i=  \mbox{min}(\delta_i, \delta_{i+1})/ \mbox{max}(\delta_i, \delta_{i+1})$, where $\delta_i= E_{i+1}- E_{i}$.
  Ergodic and localized phases are characterized by a Wigner-Dyson (WD) distributed spectrum with mean $\overline{r} \approx 0.53$ 
  and a Poisson distributed spectrum with $\overline{r} \approx 0.39$, respectively. Fig.~\ref{fig2}(a) shows the eigenstate-resolved $r$ as the disorder strength $\sigma$ varies. 
  We find a rich phase diagram featuring three distinct regimes: a critical Regime I; a nonergodic extended Regime II; and a Regime III, in which a disorder-induced flat band emerges [see Table.~\ref{table1} for the main features].  Below we shall discuss each regime in detail.

\begin{figure}
  \centering\includegraphics[width=0.48\textwidth,height=11.8cm]{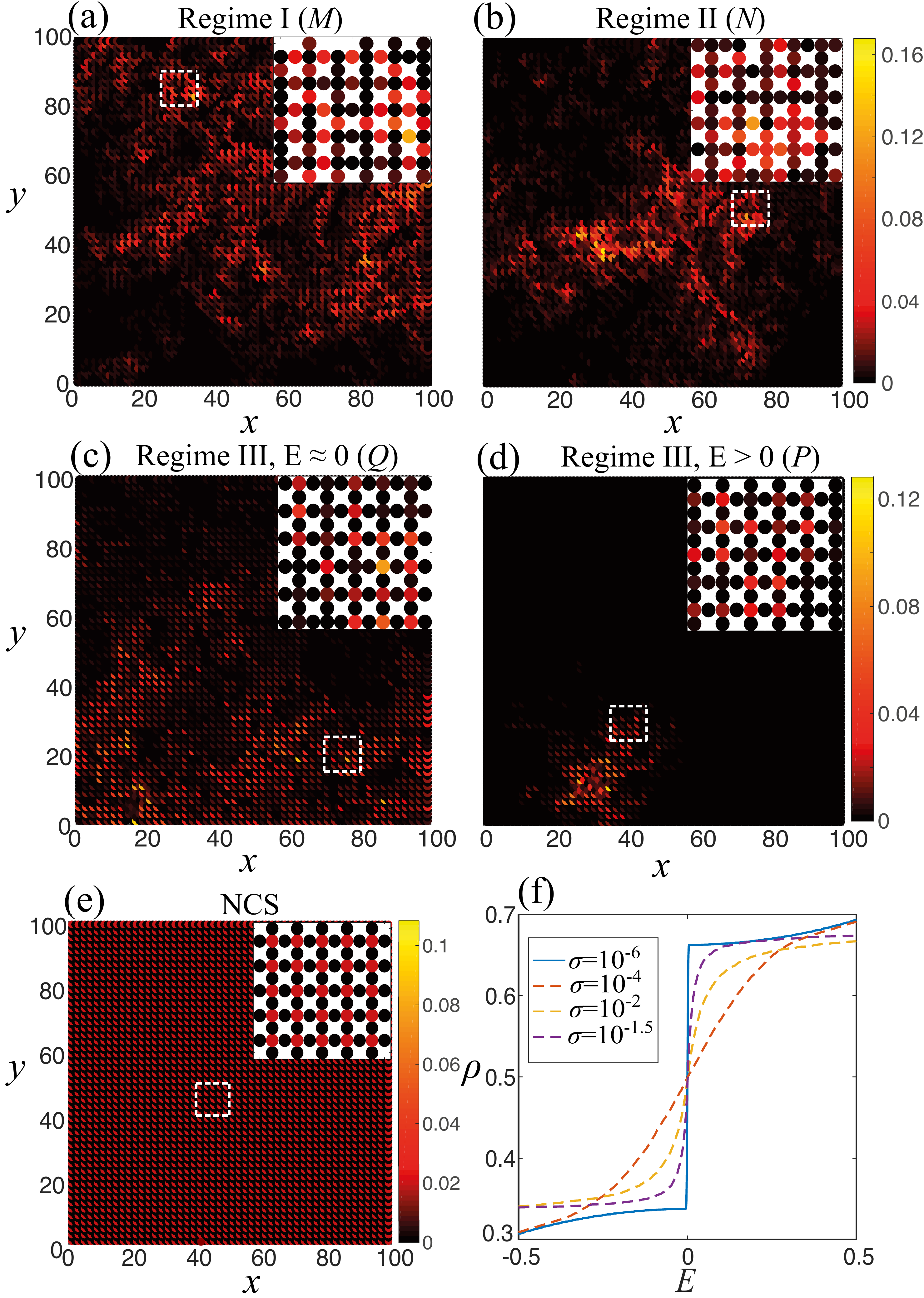}
  \caption{(a)-(d) Amplitudes of the real-space wavefunctions for representative points $M$(a), $N$(b), $Q$(c), and $P$(d) in Fig.~\ref{fig2}(a). 
  (e) The amplitudes of the wavefunction for the non-compact eigenstate (NCS). Each inset shows a zoomed-in view locally. (f) The integrated density of states as a function of energy, for different disorder strengths.
  }
\label{fig3}
\end{figure}

{\it Regime I: Criticality.---}Let us first focus on the weak-disorder regime, where the level-spacing statistics are intermediate between WD and Poisson, with the band-edge states [near the top and bottom of Fig.~\ref{fig2}(a)]  being more localized. As one can see from Fig.~\ref{fig3}(a), while the wavefunction is extended in real-space, it appears less ergodic than a perfectly delocalized state. Moreover, the wavefunction 
is mainly supported
on the $B$ and $C$ sublattices [inset of Fig.~\ref{fig3}(a)]~\cite{supp}, indicating that the flat-band states do not couple strongly to the original dispersing bands at weak disorder. 
To characterize the wavefunctions more quantitatively, we study the inverse participation ratio (IPR) $I_q(\beta) = \langle \sum_i |\psi_i^\alpha|^{2q} \rangle $, where $\psi_i^\alpha$ is the amplitude of the $\alpha$-th wavefunction on site $i$ and the average is taken over disorder realizations and over a fixed number of states $\alpha$ around $\beta$~\cite{Evers08}. It is in general expected to scale as $I_q \sim \mathcal{N}^{-{D_q(q-1)}} $, where $D_q$ is known as the fractal dimension, with $D_q=1$ for ergodic states and $D_q=0$ for localized states. If $D_q$ depends on $q$, as occurs for example at the critical point of the Anderson transition~\cite{Evers08, Luitz14, Mace19, Lindinger19, Menu20}, the eigenstates are called multifractal. Fig.~\ref{fig2}(c) shows the exponent $D_q$ extracted from the IPR for point $M$ in Fig.~\ref{fig2}(a),
which indeed exhibits a $q$ dependence, signaling multifractality and nonergodicity of the wavefunctions in this regime~\cite{Altshuler16, Luca14}.

Besides delocalization and nonergodicity  
of the wavefunctions, another interesting feature in Regime I is that the level-spacing statistics is intermediate between WD and Poisson and shows almost no dependence on system size [Fig.~\ref{fig2}(b)]. This is also clear from Fig.~\ref{fig2}(d), where we plot the 
distribution $P(s)$ of the level-spacing $s$, after spectral unfolding~\cite{Nishino07, Chalker10}, for the states shown in Fig.~\ref{fig2}(a), i.e.\ the middle one third of the states. 
This suggests that the level statistics remain intermediate between WD and Poisson in the thermodynamic limit; such statistics are called critical~\cite{Shklov93, Huang20, Chalker10, Deng16, Shukla18, Luca14}.  
The statistics also show little dependence on disorder strength, suggesting that entire Regime I is critical even for extremely weak disorder~\cite{Chalker10, Shukla18}. 
This is in contrast to the standard Anderson~\cite{Evers08} and many-body~\cite{Mace19} localization transitions, which involve a single critical point. 
The origin of the criticality in Regime I lies in the singular nature of the flat band in Hamiltonian~(\ref{eq:lieb}). As shown in Ref.~\cite{Chalker10}, for a flat band with a singular band-touching, the real-space matrix elements of the projection operator onto the flat band $\langle \mathbf R|\mathcal P|\mathbf R+\mathbf{r}\rangle$ decay as $|\mathbf{r}|^{-d}$ in $d$ dimensions. 
States originating from such flat bands are generically critical in the presence of weak disorder. On the other hand, for nonsingular flat bands (e.g.~in 1D ladder systems), $\langle \mathbf R|\mathcal P|\mathbf R+\mathbf{r}\rangle$ decays exponentially with $\mathbf{r}$ and one can use the detangling method~\cite{Marcuzzi17, Ostmann19, Baboux16} to observe  conventional Anderson localization.

{\it Regime II: Hybridization with dispersive bands.---}Similarly to Regime I, the level-spacing statistics in Regime II are also intermediate between 
WD and Poisson, as shown in Fig.~\ref{fig2}(a). However, the physics in these two Regimes is drastically different. To see this, let us first look at a representative 
real-space eigenstate in Regime II, 
shown in Fig.~\ref{fig3}(b)~\cite{supp}. Although the wavefunction is again extended but nonergodic, it now has support on all three sublattices [inset of Fig.~\ref{fig3}(b)], indicating that the original flat band strongly hybridizes with the dispersive bands as the disorder strength increases. Moreover, the fractal dimension $D_q$ again exhibits a $q$ dependence, indicating multifractality in this regime. Nonetheless, Regime II no longer appears critical, as can be seen from the noticeable but subtle system size dependence of the level statistics in Fig.~\ref{fig2}(b)~\cite{supp}.

{\it Regime III: Disorder-induced flat band.---}In the strongly disordered regime, one expects that most of the eigenstates become localized, as is indeed confirmed by the level spacing statistics in Fig.~\ref{fig2}(a). The real-space wavefunction shown in Fig.~\ref{fig3}(d) and the fractal dimension $D_q\approx 0$ in Fig.~\ref{fig2}(c) are  also consistent with the states being localized.
However, we find that in the middle of the spectrum where the energies are very close to $E= 0$, the eigenstates are \textit{delocalized} [see Fig.~\ref{fig3}(c)]. 
The fractal dimension of these delocalized states exhibits a $q$-dependence [see Fig.~\ref{fig2}(c)], indicating multifractality. Moreover, the 
integrated density of states in Fig.~\ref{fig3}(f) shows a sharper jump near $E=0$ compared to the more weakly disordered Regime II, and, counterintuitively, becomes sharper with increasing disorder. This indicates the presence of a flat band in the strong-disorder regime.
This disorder-induced flat band is physically distinct from the original flat band of Hamiltonian~(\ref{eq:lieb}) in the clean limit [solid curve in Fig.~\ref{fig3}(f)]. As can be seen from Figs.~\ref{fig3}(c)-(d), the flat-band states in the strong-disorder regime have dominant support on sublattice $A$~\cite{supp}, whereas the original flat-band states are supported on sublattices $B$ and $C$ instead [see Fig.~\ref{fig3}(a)].

To understand this disorder-induced flat band, 
we can write down the eigenvalue equation for the
single-particle hopping Hamiltonian in real space [see SM \cite{supp} for the details of the analysis in this paragraph].
By eliminating sublattice $A$ ~\cite{Hilke97}, one arrives at a single-particle hopping model on the $B$ and $C$ sublattices only, which form a \textit{planar pyrochlore lattice}. As shown in Refs.~\cite{Rhim19, Chalker10}, the planar pyrochlore lattice also hosts a singular flat band at $E=0$ in the clean limit, and the flat band eigenstates become multifractal states with $E\approx 0$ in the presence of weak disorder~[see also Fig.~\ref{fig2}(a)]. That the wavefunctions have dominant support
on sublattice $A$ in Regime III (and dominate support on $B$ and $C$ sublattices in Regime I) can also be understood using the elimination procedure.

We stress that the  disorder-induced flat band in Regime III only arises in the Rydberg atom setup, where disorder naturally couples to sublattices $B$ and $C$ only. In contrast, when disorder is present on all sublattices, as is usually the case, the density of states will instead have a broad distribution and no flat band is formed~\cite{supp}.

{\it Quench dynamics.---} The three regimes discussed above have distinct dynamical features in quantum quench experiments (see SM \cite{supp} for numerical results). We choose three different types of initial states, including a CLS, a state with nearest-neighbor Rydberg excitations, and a state with a single excitation, all of which can be prepared in  experiments~\cite{Ostmann19}. The Rydberg excitation probabilities have initial-state dependent distinct features under time evolution by the 2D disordered Lieb-lattice Hamiltonian in the three respective regimes.

{\it Conclusions and outlook.---}We have studied the effect of disorder on 2D Rydberg atom arrays in the anti-blockade regime
and uncovered rich localization phenomena depending on the disorder strength. 
In contrast to previous works~\cite{Chalker10, Leykam13, Goda06, Wilson19, Wilson20, Fu20, Hunber10, Shukla18, Gneiting2018}, our study originates from an interacting Rydberg system, and our predictions hold even in the full quantum many-body system (see SM \cite{supp}). Besides the Rydberg system, 
our results are also relevant to general disorder types \cite{supp} in other Lieb-lattice implementations, e.g., optical~\cite{Mukher15, Vicencio15, Diebel16, Guzm14, Xia16} and microwave~\cite{Kollar19} photons,
cold atoms~\cite{Taie15, Baboux16}, and 
electrons~\cite{Slot17,Jiang19}.
By changing the anti-blockade conditions, 
our study 
can be extended
to a wide variety of synthetic graphs. Moreover, our construction generically leads to
single-particle hopping models on effective graphs that are subdivisions of the graph corresponding to the physical lattice. We expect the nonergodic extended states uncovered in this work and disorder-induced flat bands to be generic for graphs with singular flat bands under this construction.
Another interesting direction is to consider 3D generalizations of our study involving the interplay of  conventional Anderson 
localization with a mobility edge and the degenerate singular bands. Finally, it would be interesting to consider 
subspaces with multiple excitations, where there can be nontrivial interplay of anti-blockade conditions and many-body interactions~\cite{kuno2020, Danieli20, Roy19} (or blockade constraints) in the synthetic lattice.

\begin{acknowledgments}
We thank Igor Boettcher, Adam Ehrenberg, Luis~Pedro Garc\'{i}a-Pintos,  Alicia Koll\'ar, Rex Lundgren, and Oles Shtanko for helpful discussions. FL, Z.-C.Y, PB, and AVG 
acknowledge funding by AFOSR, U.S.\  Department of Energy Award No.\  DE-SC0019449, AFOSR MURI, the DoE ASCR Quantum Testbed Pathfinder program (award No.\ DE-SC0019040), DoE ASCR Accelerated Research in Quantum Computing program (award No.\ DE-SC0020312),  NSF PFCQC program, ARO MURI, ARL CDQI, and NSF PFC at JQI. Z.-C.Y. is also supported by MURI ONR N00014-20-1-2325, MURI AFOSR, FA9550-19-1-0399, and Simons Foundation. TI acknowledges Iowa State University startup funds.
\end{acknowledgments}

\bibliography{localization}

\begin{thebibliography}{68}
\bibitem [{dat()}]{dataverage}%
  \BibitemOpen
  \href@noop {} {}\bibinfo {howpublished} {In (a)-(c), each point represents an
  average over eigenstates in an energy window containing 1/24 of the total
  number of eigenstates. All data shown in (a)-(d) is averaged over 800
  realizations of disorder.}\BibitemShut {Stop}%
\end{thebibliography}

\clearpage
\setcounter{figure}{0}
\makeatletter
\renewcommand{\thefigure}{S\@arabic\c@figure}
\setcounter{equation}{0} \makeatletter
\renewcommand{\thesection}{S.\Roman{section}}
\renewcommand \theequation{S\@arabic\c@equation}
\renewcommand \thetable{S\@arabic\c@table}

\begin{center} 
{\large \bf Supplemental Material}
\end{center}

This Supplemental Material consists of five sections, including the mapping from the original lattice to the Lieb lattice [\cref{sec:A}], the derivation of effective planar-pyrochlore hopping model [\cref{sec:I}], level spacing statistics in Regime II [\cref{sec:II}], sublattice-resolved wavefunction weight distributions in each regime [\cref{sec:III}], additional numerical results on uncorrelated disorder as well as disorder that couples to all sublattices [\cref{sec:IV}], numerical results on the quench dynamics in three regimes [\cref{sec:V}], and numerical results for the full quantum many-body system [\cref{sec:VI}].

\section{Mapping from the original lattice to the Lieb lattice }

\label{sec:A}

In this section, we illustrate the mapping from the original square lattice to the synthetic Hilbert space lattice (Lieb lattice in Fig.~1 in the main text or Fig.~\ref{figs1}). As described in the main text, the anti-blockade (facilitation) condition is obtained by setting $\Delta= V(R_0)$, so that an isolated excitation makes the excitation of its nearest neighbor resonant~\cite{Mattio15, Marcuzzi17, Ostmann19}. We also
work in the limit
$ |\Delta| \gg \Omega$,   
where the un-facilitated excitations are sufficiently off-resonant. Additionally, it is required that $V(\sqrt{2} R_0), V(2R_0) \gg \Omega$, so that a pair of neighboring Rydberg excitations is unable to further facilitate the excitation of any neighboring site.

\begin{figure}[ht]
  \centering\includegraphics[width=0.36\textwidth]{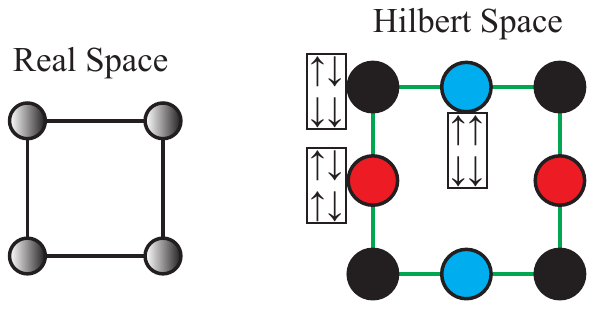}
  \caption{ The mapping of local site configuration between the original square lattice and the synthetic Hilbert space lattice (Lieb lattice). In the real space configuration, each dot denotes the position of a trapped tweezer. In the Hilbert space configuration, the black dots denote single Rydberg excitation, while the red and blue dots denote pairs of neighboring excitations. 
  }
\label{figsa}
\end{figure}

Under these conditions, the Hilbert space is split into subspaces that are separated by an energy scale much larger than $\Omega$. Each subspace contains 'quasi-resonant' 
states separated by an energy $ \sim O(\Omega)$. Intuitively, under the action of Hamiltonian (1) in the main text, an isolated excitation can at most produce one more in its neighborhood, after which either the former facilitates the de-excitation of the latter, or vice versa (see Fig. \ref{figsa}). In the $x$ (or $y$) direction, such process can be described by
\begin{equation}
    \ket{...\downarrow \uparrow \downarrow \downarrow...} \longleftrightarrow \ket{...\downarrow \uparrow \uparrow \downarrow...} \longleftrightarrow \ket{...\downarrow \downarrow \uparrow \downarrow...} . 
    \label{eq_map}
\end{equation}
If we use black dots to denote single excitations and blue (red) dots to denote pairs of neighboring excitations in the $x$ ($y$) direction, the connectivity given by Eq.~(\ref{eq_map}) is described by a Lieb lattice (see Fig.~1 in the main text or Fig.~\ref{figs1}). Note that the blue and red dots are not directly connected in the Lieb lattice graph, because the hopping between the two sites is a second-order process and the matrix element between the two states vanishes. 

As discussed in the main text, the Lieb lattice supports flat band states. There are mainly three types of flat band states in a Lieb lattice without disorder. The compact localized states (CLSs) are states localized in real space. The non-contractible loop states (NLSs) are states with support that is localized in one direction but extended in the other, as shown in Fig.~1(b) in the main text. It turns out these states are essential for the critical nature of the states in Regime I (see main text for more details). The third type of flat band states is the non-compact state (NCS), which is extended in real space in both $x$ and $y$ directions. This is illustrated in Fig.~1(c) in the main text.

\section{Derivation of effective planar-pyrochlore hopping model}
\label{sec:I}

\begin{figure}[ht]
  \centering\includegraphics[width=0.45\textwidth]{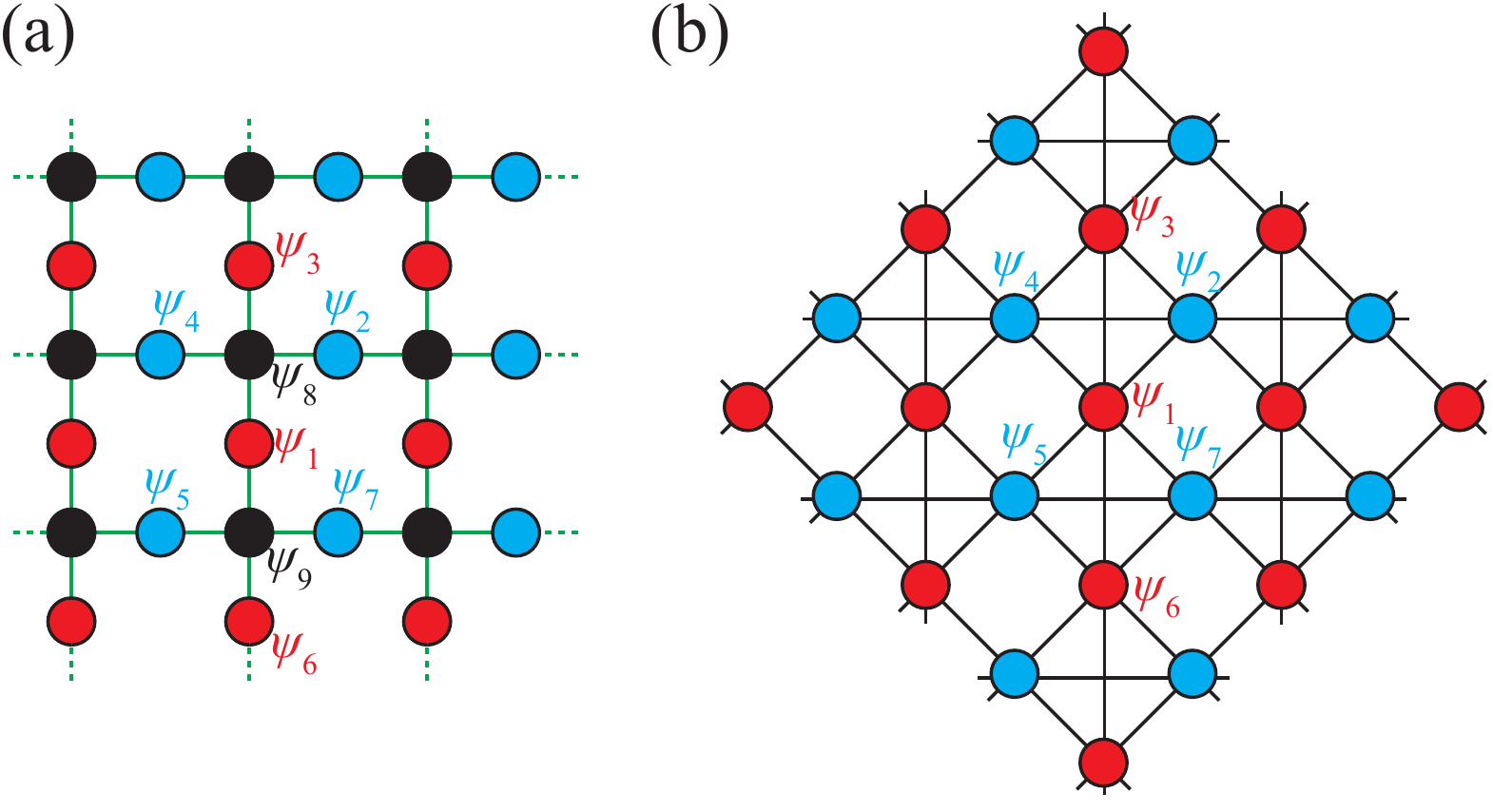}
  \caption{ (a) The Lieb lattice and the wavefunction amplitudes $\psi_1, \ldots, \psi_9$ in a local region. (b) The mapped planar-pyrochlore hopping model after eliminating the sublattice $A$. 
  }
\label{figs1}
\end{figure}

In this section, we derive the effective planar-pyrochlore hopping model by eliminating the wavefunction amplitudes on sublattice $A$. For clarity, we  show the Lieb lattice and the wavefunction amplitudes $\psi_1, \ldots, \psi_9$ on each lattice site in Fig.~\ref{figs1}. 

Let us focus on  a particular eigenstate with eigenenergy $E$. The eigenvalue equation centered around the site with wavefunction amplitude $\psi_1$ is given by
\begin{equation}
   \frac{\Omega}{2} (\psi_8 +\psi_9) + V_1 \psi_1= E \psi_1, 
   \label{eqs1}
\end{equation}
where $V_1$ is the disorder strength on site 1. Similarly, we can write down the eigenvalue equations for $\psi_8$ and $\psi_9$: 
\begin{subequations}
\label{eqs2}
\begin{align}
\frac{\Omega}{2} (\psi_1 +\psi_2+ \psi_3+ \psi_4)  = E \psi_8, \\
\frac{\Omega}{2} (\psi_1 +\psi_5+ \psi_6+ \psi_7)  = E \psi_9. 
\end{align}
\end{subequations}
By plugging Eqs.~(\ref{eqs2}) into Eq.~(\ref{eqs1}) and multiplying through by $4E$, one arrives at~\cite{Hilke97}:
\begin{equation}
\Omega^2 \sum_{i=2}^7 \psi_i + (4 EV_1+2\Omega^2)\psi_1 = 4 E^2 \psi_1,
\label{couple}
\end{equation}
where $V_1$ is the on-site random potential on site 1. The above equation corresponds to a single particle hopping on the planar pyrochlore lattice, see Fig.~\ref{figs1}(b).
Eq.~(\ref{couple}) now describes a single-particle hopping model
on the $B$ and $C$ sublattices only, which form a \textit{planar pyrochlore lattice}. As shown in Refs.~\cite{Rhim19, Chalker10}, the planar pyrochlore lattice also hosts a singular flat band in the clean limit, and the flat band eigenstates also exhibit multifractality in the presence of weak disorder. Indeed, for $E\approx 0$, the right-hand side of Eq.~(\ref{couple}) can be neglected, and when $EV_1 \ll \Omega^2$, Eq.~(\ref{couple}) describes a single-particle hopping 
model in the presence of weak disorder. That the wavefunctions have dominant support
on sublattice $A$ can also be understood. When eliminating sublattice $A$ from the eigenvalue equations, we have used $\psi_8=\frac{\Omega}{2 E}(\psi_1+\psi_2+\psi_3+\psi_4)$, where site 8 belongs to sublattice $A$ [see Fig.~1(b)]. When $E\approx 0$, the weight on sublattice $A$ is enhanced. On the other hand, in Regime I near the clean limit, the original CLS has $\psi_1+\psi_2+\psi_3+\psi_4=0$, hence the weight on sublattice $A$ remains negligible.

\begin{figure}
  \centering\includegraphics[width=0.36\textwidth]{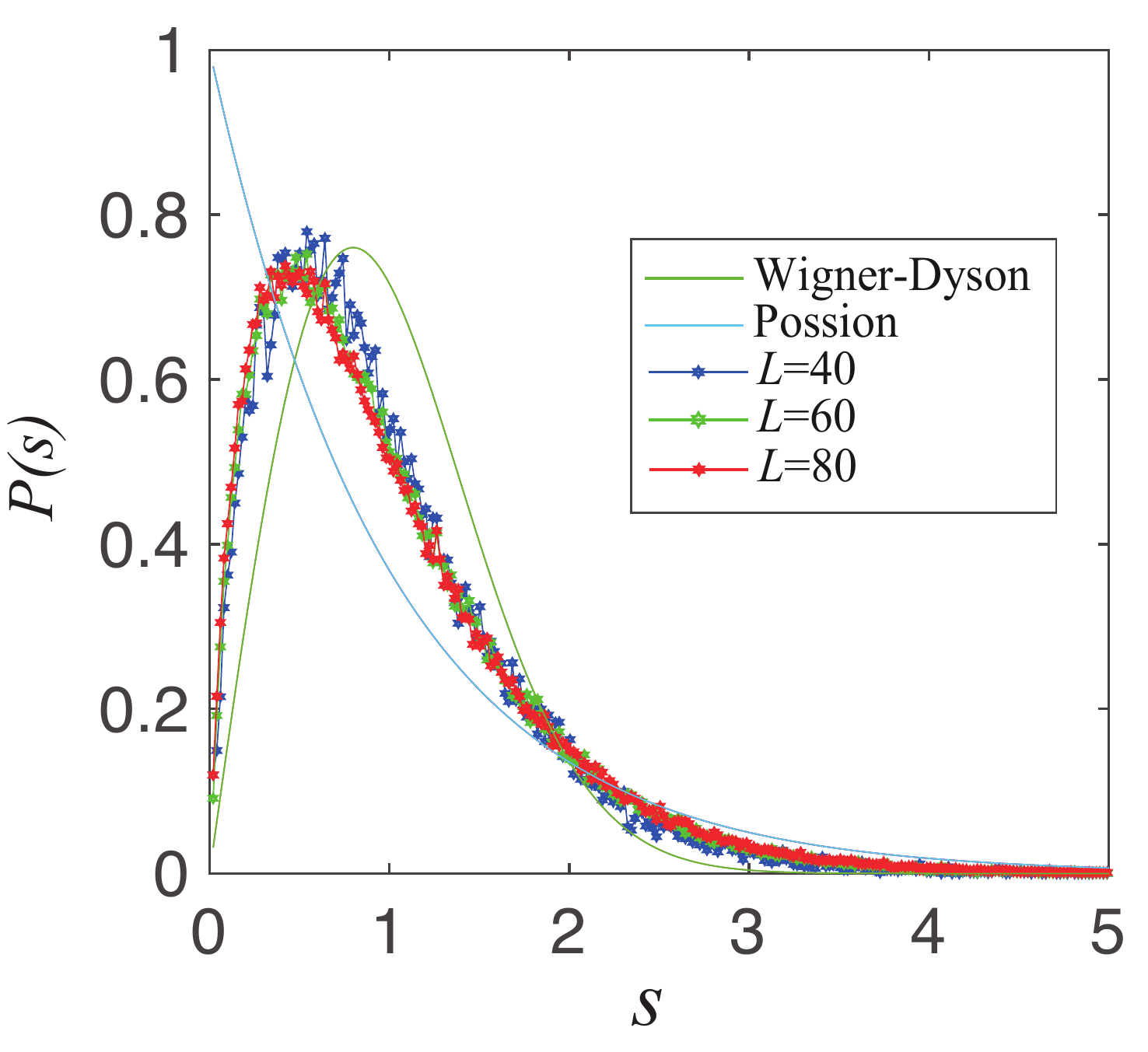}
  \caption{ Probability distribution of the unfolded level spacings $P(s)$ for states in Regime II for different system sizes. The disorder strength $\sigma= 10^{-3.6}$. The data shown in this figure are averaged over 800 realizations of disorder.
  }
\label{figs2}
\end{figure}

\section{Level spacing statistics in Regime II}
\label{sec:II}

In Fig.~\ref{figs2}, we show the probability distribution of the unfolded level spacings $P(s)$ for states in Regime II for different system sizes. Compared to that in Regime I [Fig.~2(d)], we find that the system-size dependence of the level spacing distribution in Regime II is more prominent.
The level statistics appear to tend towards Poisson as the system size increases. This suggests that Regime II is not critical, but rather localized in the thermodynamic limit.

\section{Sublattice-resolved wavefunction weight}
\label{sec:III}

\begin{figure}
  \centering\includegraphics[width=0.5\textwidth]{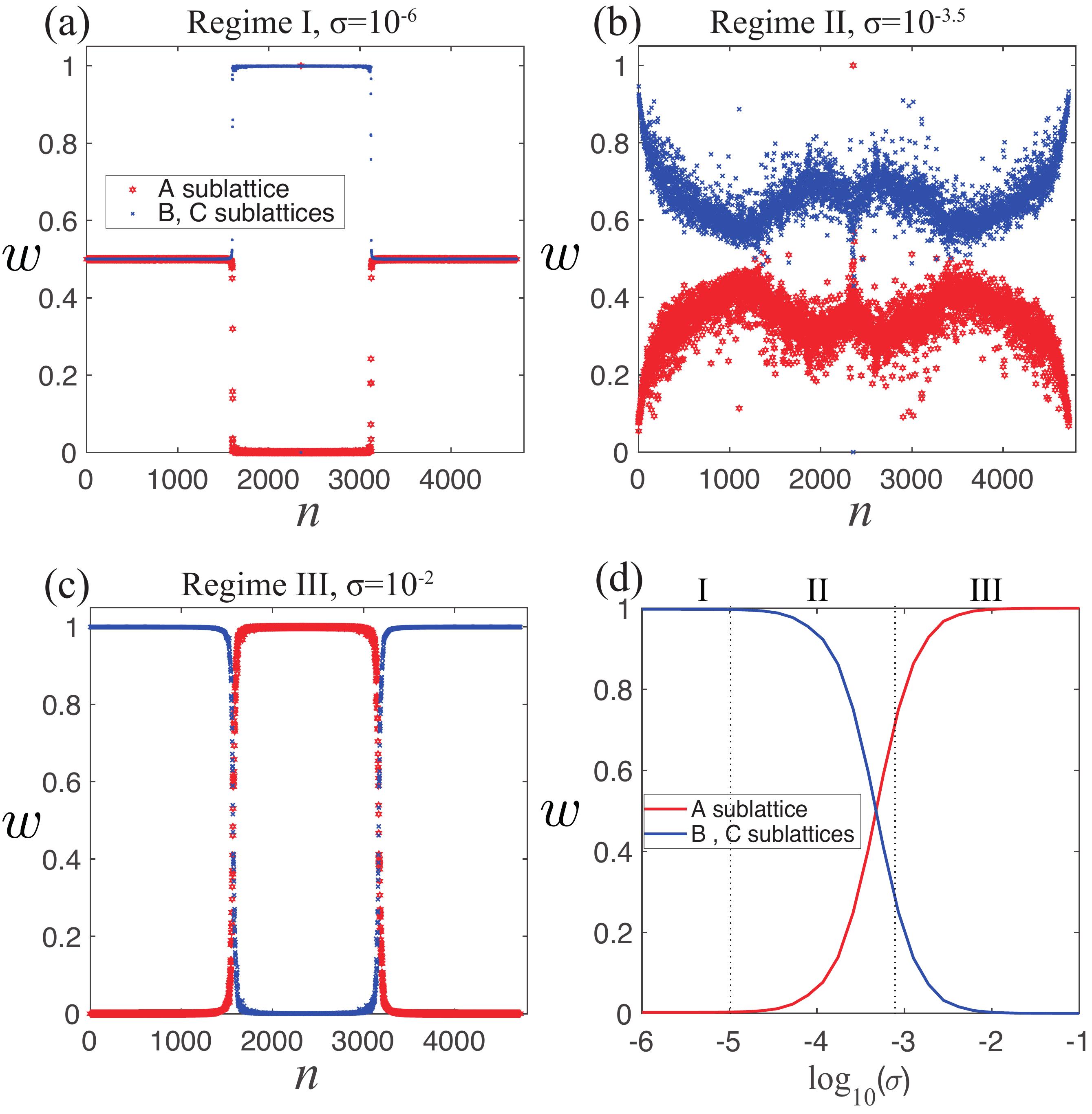}
  \caption{ (a)-(c) Sublattice dependence of the wavefunction weight $w$ versus the eigenstate number $n$ sorted by eigenenergy in each regime.  (d) $w$ versus the disorder strength for states near $\beta=0$ (averaged over 1/24 of the total number of eigenstates). The system size is $L=40$.  The dashed vertical lines separate the three Regimes, as in Fig.~2(a) of the main text. 
  }
\label{figs3}
\end{figure}

In this section, we plot the sublattice dependence of the wavefunction weight in each regime. From Fig.~\ref{figs3}(a), one can see that, in Regime I, the wavefunction weights $w$ in the middle one-third of the spectrum have major support on the $B$ and $C$ sublattices, and negligible support on the $A$ sublattice. This is consistent with our discussion in the main text that disorder in this regime only slightly modifies the flat band in the clean limit, and does not induce hybridization with other bands.
In contrast, the wavefunction in Regime II has 
substantial support on all lattice sites [Fig.~\ref{figs3}(b)]. This indicates that disorder in Regime II induces strong hybridization between the flat band and the dispersive bands. In Regime III, strong disorder gives rise to a new flat band in the middle of the spectrum, whose wavefunctions now have major support on sublattice $A$, in stark contrast to the flat band in Regime I. 
Another interesting observation in this Regime is that eigenstates 
in the rest of the spectrum show opposite support, i.e.~they are mainly supported on the $B$ and $C$ sublattices. These eigenstates do not belong to the flat band, and the corresponding energies are far from zero.

\begin{figure} 
  \centering\includegraphics[width=0.5\textwidth]{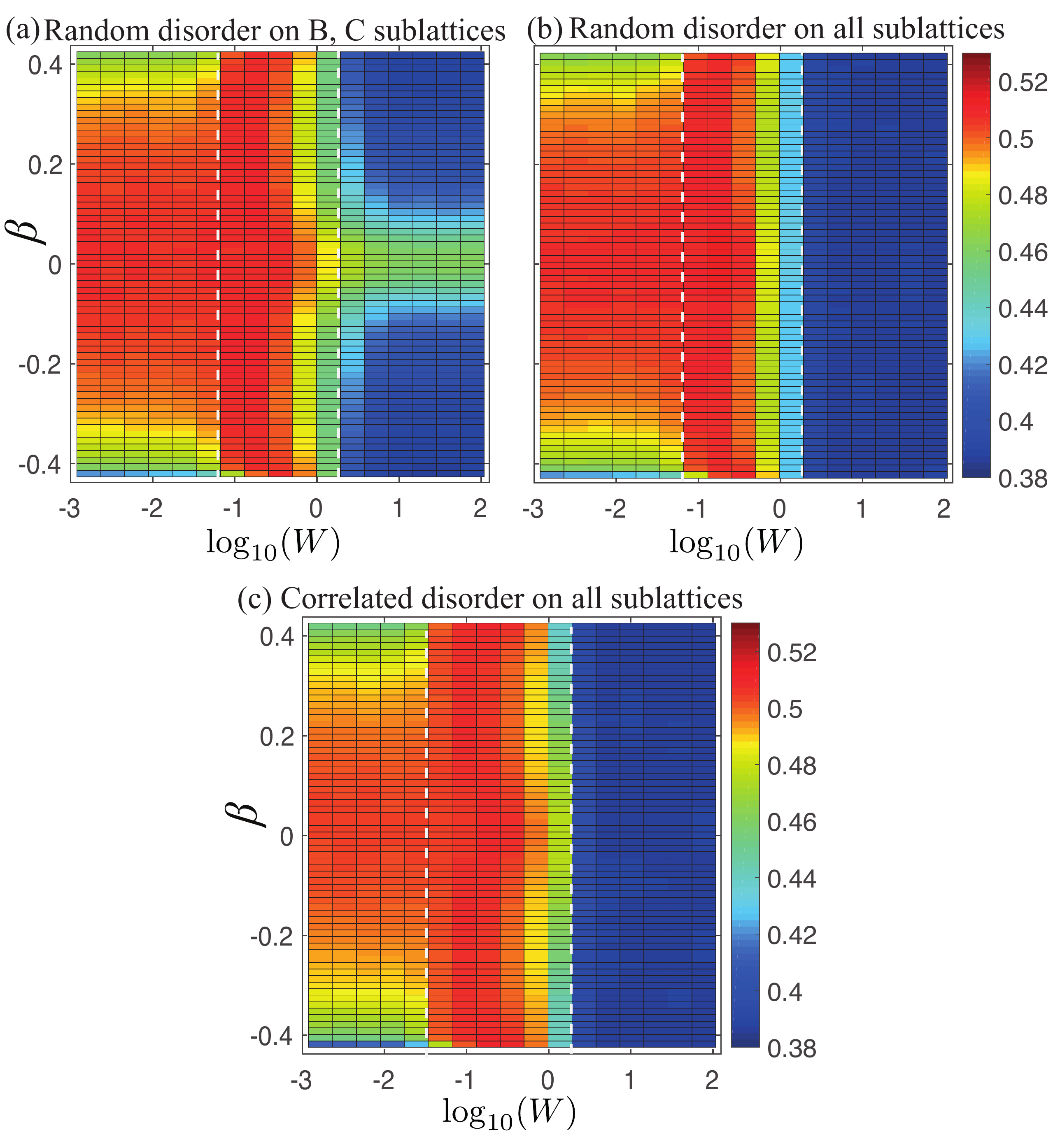}
  \caption{ (a) Level spacing ratio $r$ for uncorrelated disorder coupled to the $B$ and $C$ sublattices. (b) Level spacing ratio $r$ for random disorder coupled to all sublattices. (c) Level spacing ratio $r$ for correlated disorder on all sublattices in which the shift on $B$ and $C$ sites is given by the sum of the shifts on the two neighbouring $A$ sites. The disorders in (a) and (b) both have a uniform distribution in $[-W, W]$. The disorder in $(c)$ uses the same uniform distribution for the $A$ sites only. The system size is $L=60$. The data shown in this figure are averaged over 800 realizations of disorder.
  }
\label{figs4}
\end{figure}

Fig.~\ref{figs3}(d) shows the wavefunction weight as a function of the disorder strength for states near $\beta=0$. From this figure, one can see that there is indeed a flip of the support from the $B$ and $C$ sublattices to the $A$ sublattice as the disorder strength increases.
This plot also corroborates the existence of three distinct localization regimes as discussed in the main text.
\\

\begin{figure*}
  \centering\includegraphics[width=0.92\textwidth, height=15cm]{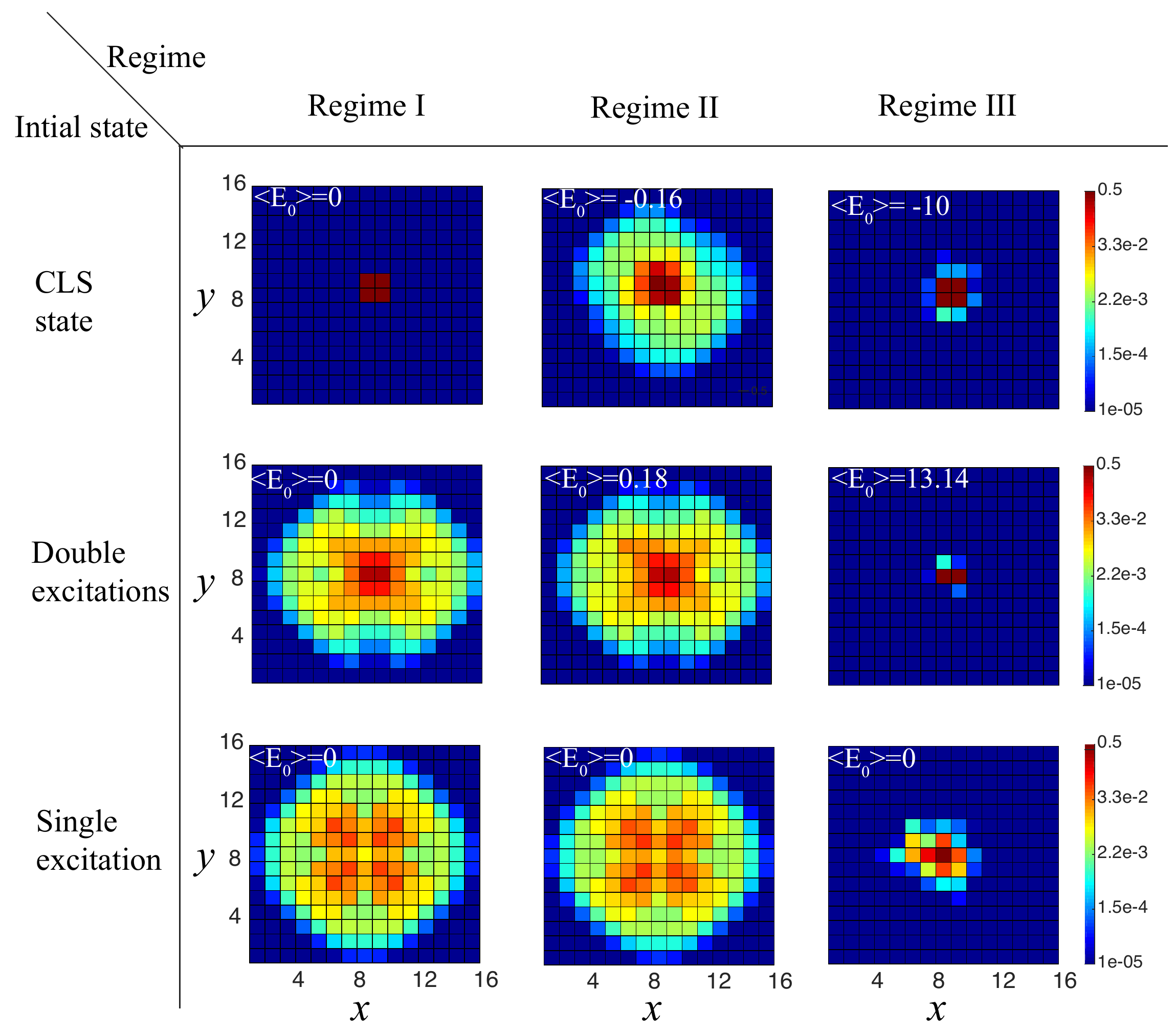}
  \caption{ Rydberg excitation probability in real space after evolving for a time  $10/\Omega$ under the 
  2D Lieb-lattice Hamiltonian with disorder strengths $\sigma=10^{-6}$ (first column),  $\sigma=10^{-4}$ (second column),  and $\sigma=10^{-2}$ (third column). $V= 200 \Omega$ is chosen for all plots.  Three types of initial states are considered: CLS (first row), double-excitation (second row), and single excitation (third row). $\langle E_0 \rangle$ denotes the single-particle energy expectation value for each initial state.
  }
\label{figs5}
\end{figure*}

\section{Numerical results on different disorder types}
\label{sec:IV}
In the main text, we have assumed that the $A$ sublattice sites do not couple to the disorder. Here we validate this approximation. First, suppose that we do not have a magic lattice. The differential AC Stark shift experienced by
Rydberg atoms relative to ground-state atoms indeed could then lead to on-site disorder coupled to the A sublattice. However, this type of disorder is much smaller compared to the disorder due to interactions. For instance, in Regime III with the largest positional disorder, the change in the differential AC Stark shift due to disorder is around $1.7$ MHz, where we chose the following experimentally relevant parameters: $1$ mK trap depth, $0.9 \mu$m waist, $3.76 \mu$m lattice constant, 300 MHz nearest-neighbour interacting strength,  and disorder strength $\sigma= 0.01$. This disorder arising from the AC Stark shift is much smaller than the $\sim 10$ MHz disorder arising from the interaction in the same regime. We have also checked that in Regimes I and II, the disorder caused by the AC Stark shift is also much weaker than the interaction-induced disorder.

On the other hand, if we assume a magic lattice, the ground-state and Rydberg-state atoms will experience the same potential. In this case, the detuning  will be homogeneous in space. Moreover, even if different atoms are in different motional states, the detuning will still be the same for all atoms if they do not change their motional state as they make a transition between ground and Rydberg states. This, in turn, can be achieved by making sure that the excitation lasers and hence the Rabi frequency $\Omega$ are spatially uniform. In such a case, the on-site disorder on the A sublattice will vanish.

To be more comprehensive, we  carry out numerical studies of more general types of disorder in this section. We provide numerical results on three other types of disorder as shown in Figs.~\ref{figs4}(a)-(c), namely, (a) uncorrelated disorder on sublattices $B$ and $C$ only, (b) uncorrelated disorder coupled to all three sublattices, and (c) correlated disorder on all sublattices in which the shift on $B$ and $C$ sites is given by the sum of the shifts on the two neighbouring $A$ sites.
Disorder of type (b) can be implemented in other experimental platforms, e.g., lattice systems of optical photons, microwave photons, cold atoms, and electrons. Disorder of type (c) corresponds to the frequency shift due to the Rydberg setup's Doppler effect. While this effect is present in the Rydberg platform discussed in the main text, we neglect it because the positional disorder is much stronger~\cite{Bernien17}.

We first consider uncorrelated disorder coupled to the $B$ and $C$ sublattices only with a uniform distribution in $[-W, W]$, in contrast to the correlated disorder studied in the main text.  From Fig.~\ref{figs4}(a), we again find three regimes similar to Fig.~2(a) in the main text. 
We have also confirmed that the properties of the three regimes' wavefunctions are similar to those of the correlated-disorder case. We thus conclude that the presence or absence of correlations in the disorder on adjacent lattice sites has little effect on the main results of this work.

Fig.~\ref{figs4}(b) and (c) also show three regimes. However, in stark contrast to Fig.~\ref{figs4}(a) and Fig.~2(a) in the main text, all states in Regime III are strongly localized, and the flat band with extended states near the middle of the spectrum is absent.
This is because the disorder is present on all sites, and all states in this regime undergo an Anderson localization transition. 
On the other hand, we have checked that the wavefunction properties in Regimes I and II are similar to the case where disorder couples to the $B$ and $C$ sublattices only.

\section{Numerical results for quench dynamics \label{sec:V}}

In this section, we provide numerical results demonstrating that the three regimes discussed in the main text have distinct dynamical features in quantum quench experiments. We choose three different types of initial states, including a CLS, a state with nearest-neighbor Rydberg excitations, and a state with a single excitation [see Fig.~\ref{figs5}].
The CLS corresponds to $\ket{\Psi_{ij}}=\frac{1}{2}\left( \ket{11}_{i,j }^x +   \ket{11}_{i, j+1 }^x-\ket{11}_{i,j }^y- \ket{11}_{i+1,j }^y \right) $,  where $ \ket{11}_{i,j }^{x/y}$ represents a pair of nearest-neighbour Rydberg excitations emanating from site $(i, j)$ in the $x/y$ direction. All the other atoms are in the ground state. Such a state can be experimentally prepared with single-site addressing. The main idea is first to use pulses with blockade conditions to prepare a single excitation and then use pulses with anti-blockade conditions to prepare pairs of nearest-neighbor excitations. 
The reader can find a detailed pulse schedule in Ref.~\citep{Ostmann19}. Here, we also consider initial states with single and nearest-neighbor Rydberg excitations, which can be prepared using fewer pulses than the CLS.

Fig.~\ref{figs5} shows the Rydberg excitation probabilities in real-space after evolving for a fixed amount of time under the
2D disordered Lieb-lattice Hamiltonian in the three respective regimes. The total evolution time is chosen to be on the scale of microseconds, in which case the loss of Rydberg atoms is negligible \cite{Bernien17}. In Regime I 
(weak disorder), the CLS initial state hybridizes weakly with
other flat-band states, and hence the distribution of Rydberg excitations spreads slowly in time. Within the experimentally accessible timescale, we observe hardly any spreading of the Rydberg excitations. However, if we start with initial states with a single excitation or a nearest-neighbor excitation, the Rydberg excitations spread quickly. This is due to the fact that the single- and double-excitation initial states hybridize strongly with the dispersive bands.  In Regime II, the initial state couples to both the flat-band states and dispersive bands. The Rydberg excitations thus spread much faster in this case for all three types of initial states. Finally, in the strong-disorder Regime III, the CLS and double-excitation initial states strongly couple to localized states far from the flat band, and the Rydberg excitations are strongly localized around their initial positions. However, when the initial state only contains a single excitation, it couples to the disorder-induced flat band at $E=0$. The excitation shows non-ergodic but delocalized dynamics.

In this work, we are assuming the frozen-gas limit. Let us now confirm the validity of this assumption. 
Typical atomic velocity $v$ and disorder strength $\sigma$ (in units of lattice constant $R_0$)
are related to the temperature $T$ via $mv^2 \sim m \omega^2 (R_0 \sigma)^2 \sim k_B T$, where $m$ is the atom mass and 
$\omega$ is the trapping frequency.
During the evolution after a quantum quench, the distance an atom shifts over time $t$ compared to the original disorder length $\sigma R_0$ is $\frac{\delta x}{ R_0 \sigma} \sim \frac{v t}{R_0 \sigma} \sim \omega t$.
Using a typical trapping frequency ($\omega \approx (2 \pi) 10$ kHz) and evolution time ($t \approx 1 \mu$s), we have $\omega t \approx 0.06 \ll 1$,
confirming the validity of the frozen-gas approximation.\\

\begin{figure}
  \centering\includegraphics[width=0.52\textwidth, height=9.8cm]{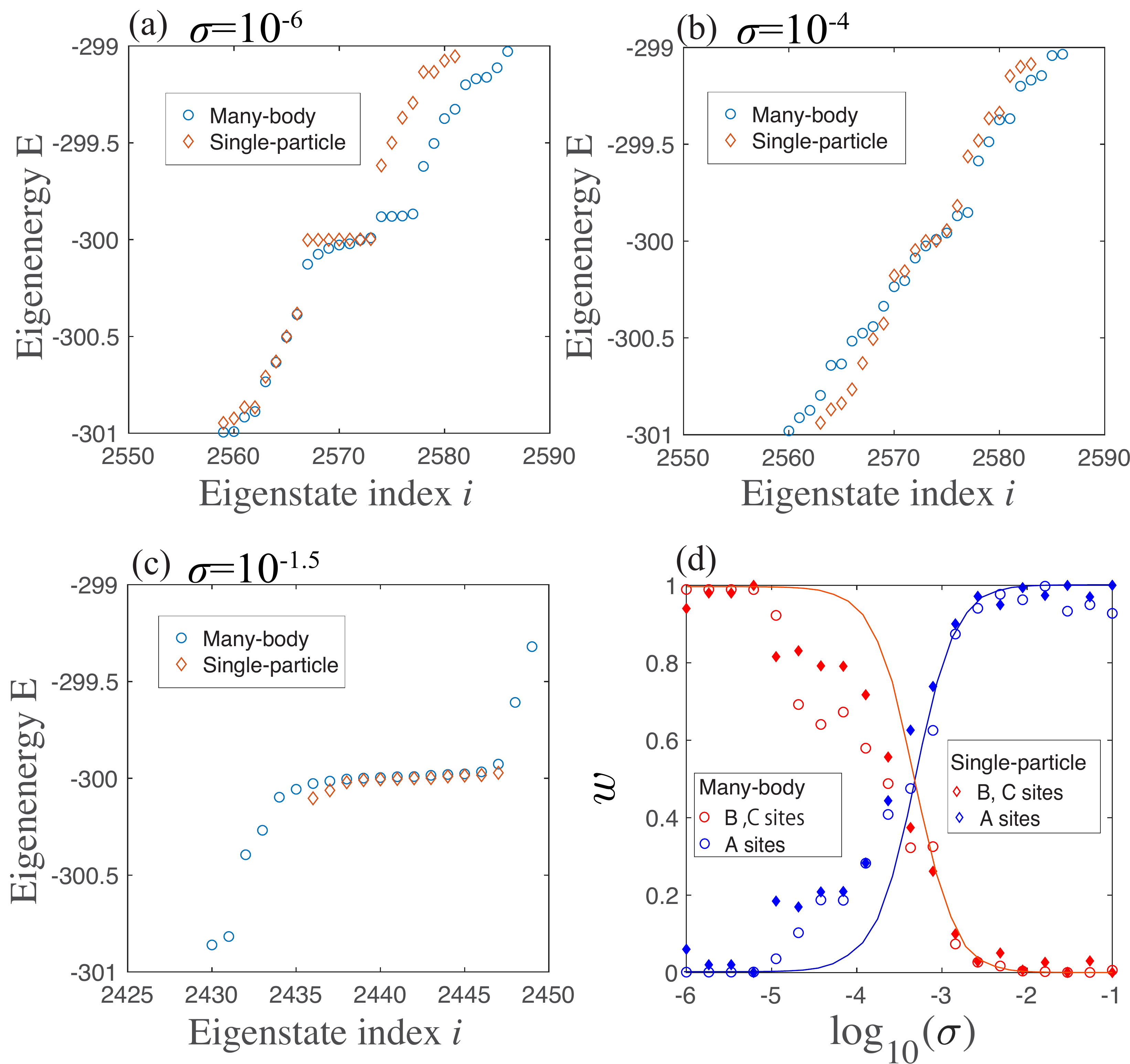}
  \caption{  Many-body and Single-particle energy spectrum near eigen-energy $E= -V_0$ for disorder strength (a) $\sigma= 10^{-6}$ (b) $\sigma= 10^{-4}$  and (c) $\sigma= 10^{-1.5}$.  (d) Weights of many-body (circles) and single-particle (diamonds) states on A or B and C lattices versus disorder strength. All the results are obtained for a $4\times3$ lattice and averaged over 6 states near $E= -V_0$.  The lines are the results from the projected single-particle Hamiltonians with system size $40\times40$ [same as Fig.~\ref{figs3}(d)]. 
  }
\label{figs6}
\end{figure}

\section{Numerical results for the full quantum many-body system }
\label{sec:VI}

In this section, we carry out full numerical calculations for the quantum many-body  system on a  $4 \times 3$ lattice and illustrate how the features of quantum states in the full many-body case correspond to the single-particle results presented in the main text.

We first study the spectrum of many-body eigenstates near $E= -V_0$.  As shown in Fig.~3(f) in the main text, with increasing disorder, the single-particle  spectra (calculated within the projected subspace associated with the facilitation conditions) first becomes dispersive and then becomes flat again (in the strong-disorder limit).  In Figs.~\ref{figs6}(a)-(c), we compare the many-body and single-particle energy spectrum near the flat band energy $E= -V_0$. We find that the many-body spectrum also shows highly consistent features. In addition, we see that the disorder-induced flat band is also a concrete feature of the full many-body problem [Fig.~\ref{figs6}(c)]. 

We also study the projection of many-body states onto the reduced subspace. As shown in Fig. \ref{figs3}(d), when the disorder is small, the single-particle wavefunctions are supported mainly on the B and C sublattices (corresponding to many-body states with pairs of neighboring excitations); when the disorder is large, the single-particle wavefunctions are mainly supported on the A sublattice (corresponding to many-body states with single excitations). In Fig.~\ref{figs6}(d), we project many-body wavefunctions with energy near $-V_0$ into these subspaces. It is clear that the many-body wavefunctions are mainly composed of states from the appropriate sublattices. The transition of support from the B and C sublattices to the A sublattice as disorder increases is also a clearly visible feature of the many-body wavefunctions. For comparison, we also plot the wavefunction weight calculated from the single-particle Hamiltonian for a $4 \times 3$ lattice. Again, we see the consistent features between many-body and single-particle systems even for such a small system size (where attribute the deviations to finite-size effect).

In summary, although we are limited by system size for the full many-body Hamiltonian, the above comparisons show strong evidence that our approximation is valid, and all the interesting results in our paper hold for the full many-body system.

\end{document}